\documentclass[structabstract]{aa}  
\usepackage{graphicx}
\usepackage{txfonts}
\usepackage{natbib}

\begin{document}
   \title{Properties of starspots on CoRoT-2\thanks{CoRoT is a space project operated by the French Space Agency, CNES, with participation of the Science Programme of ESA, ESTEC/RSSD, Austria, Belgium, Brazil, Germany, and Spain.} 
}

   \subtitle{}

   \author{Adriana Silva-Valio
          \inst{1}
          \and
          A. F. Lanza\inst{2} \and R. Alonso\inst{3} \and P. Barge\inst{3}
          }

   \institute{CRAAM, Mackenzie University, Rua da Consola\c c\~ao, 896, 01302-907, S\~ao Paulo, Brazil\\
              \email{avalio@craam.mackenzie.br}
         \and INAF-Osservatorio Astrofisico di Catania, Via S. Sofia, 78, 95123 Catania, Italy            
         \and Laboratoire d'Astrophysique de Marseille, UMR 6110, CNRS/Universit´e de Provence, Traverse du Siphon, 13376 Marseille, France
             }

   \date{Received ; accepted }

 
  \abstract
   {As a planet eclipses its parent star, a dark spot on the surface of the star may be occulted, causing a detectable variation in the light curve. }
   {Study these light curve variations during  transits and infer the physical characteristics of the stellar spots.}
   {A total of 77 consecutive transit light curves of CoRoT-2 were observed with a high temporal resolution of 32 s, corresponding to an uninterrupted period of 134 days.  By analyzing small intensity variations in the transit light curves, it was possible to detect and characterize spots at fixed positions (latitude and longitude) on the surface of the star. The model used simulates planetary transits and enables the inclusion of spots on the stellar surface with different sizes, intensities ({\it i.e.} temperatures), and positions. Fitting the data by this model, it is possible to infer the spots physical characteristics. Because what is observed is the stellar flux blocked by the spots, there is a degeneracy between the spots intensity and area. Thus the fits were either in spot longitude and radius, with a fixed intensity, or in spots longitude and intensity, for spots of constant size. The model allowed up to 9 spots to be present at the stellar surface within the transit band. }
   {Before the modeling of the spots were performed, the planetary radius relative to the star radius was estimated by fitting the deepest transit to minimize the effect of spots. A slightly larger (3\%) radius, 0.172 $R_{star}$, resulted instead of the previously reported 0.1667 $R_{star}$. The fitting of the transits yield spots, or spot groups, with sizes of ranging from 0.2 to 0.7 planet radius, $R_p$, with a mean of $0.41 \pm 0.13 R_p$ ($\sim$ 100,000 km), resulting in a stellar area covered by spots within the transit latitudes of 10-20\%. The intensity varied from 0.4 to 0.9 of the disk center intensity, $I_c$, with a mean of $0.60 \pm 0.19 I_c$, which can be converted to temperature by assuming blackbody emission for both the photosphere and the spots. Considering an effective temperature of 5625 K for the stellar photosphere, the spots temperature ranges mainly from 3600 to 5000 K.  }
   {The spot model used here was able to estimate the physical characteristics of the spots on CoRoT-2, such as size and intensity. These results are in agreement with those found for magnetic activity analysis from out of transit data of the same star.  }


   \maketitle
%

\section{Introduction}

Exactly four centuries ago, spots on the surface of the Sun were first detected by Galileo. Sunspots are cool regions of strong concentration of magnetic fields on the photosphere of the Sun. Presently, thanks to the observations with unprecedented photometric precision of the CoRoT satellite \citep{baglin06}, spots on the surface of another star can be detected, that of CoRoT-2, a much more active star. Such activity is probably a consequence of its young age, estimated in 0.5 Gyr \citep{bouchy08}. CoRoT-2 is one of the 7 stars with transiting planets detected so far with CoRoT.

Spot activity has been inferred in the past from the modulation observed in the light of stars. As the star rotates, different spots on its surface are brought into view. Because spots are cooler, and therefore darker, than the surrounding photosphere, the detected total light of the star diminishes as different spots of varying temperature and size face the observer. This periodic modulation enables the determination of the rotation period of the star.

The magnetic activity of CoRoT-2 star has been studied in detail by \cite{lanza09} who analyzed its out of transit light curve with modulations of $\sim$6\% of its total flux. Using maximum entropy regularized models, Lanza and collaborators (2009) modeled the light curve considering both the presence of sunspots and faculae (dark and bright regions, respectively). This study detected the existence of two active longitudes located in opposite hemispheres, and also that these longitudes varied with time. The active longitudes are regions where the spots preferably form. The total area covered by the spots were seen to vary periodically with a period of $29 \pm 4$ days. The authors were also able to estimate an amplitude for the relative differential rotation of $\leq 0.7$\%. 

Here we propose to study these same spots, however, with a different approach. A transiting planet can be used as a probe of the contrasting features in the photosphere of the star. When the planet occults a dark spot on the surface of its host star, small variations in the light curve may be detected \citep{silva03, pont07, silva-valio08}. From modeling of these variations, the properties of the starspots can be determined, such as size, position, and temperature. Moreover, from the continuous and long duration of observation provided by the CoRoT satellite, the temporal evolution of individual spots can be obtained.

The next session describes the observation performed by CoRoT, whereas the model used here is introduced in the following session. The main results are presented in Session 4. Finally, a discussion of the main results and the conclusions are listed in Session 5.    


\section{Observation of CoRoT-2}

A planet around the star CoRoT-2 was detected during one of the long run observations of a field toward the Galactic center performed by the CoRoT satellite, and is the second transiting planet discovered. The planet, a hot Jupiter, with a mass of $3.3 M_{Jup}$ and radius of $1.47 R_{Jup}$, orbits its host star in just 1.73 day \citep{alonso08}. CoRoT-2 is a solar-like star of type G7, with 0.97 $M_\odot$ and 0.902  $R_\odot$ that rotates with a period of 4.54 days \citep{alonso08}. The parameters of the star and the planet plus the orbital ones were determined from a combination of CoRoT photometric light curve \citep{alonso08} and ground based spectroscopy \citep{bouchy08}. 

The data analyzed here were reduced following the same procedure as \cite{alonso08}.
The light curve was filtered from cosmic ray impacts and orbital residuals. Then this cleaned light curve was folded considering an orbital period of 1.743 day, from which the parameters of the planetary system were derived. Beside the planet orbital period of $P_o = 1.743$ day and stellar rotation of $P_s = 4.54$ days, the orbital parameters were: semi-major axis of $a = 6.7$ star radius ($R_s$) and inclination angle of $87.84^\circ$.

\begin{figure}
   \centering
  \includegraphics[width=8cm]{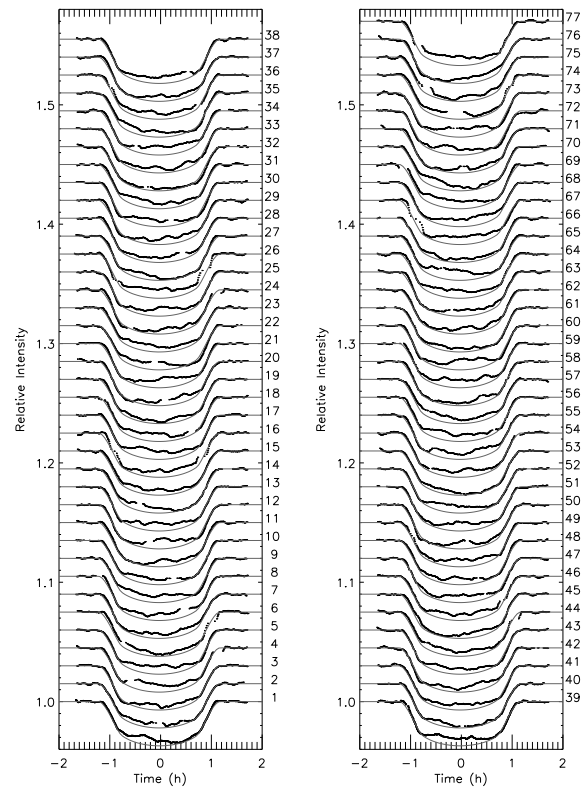}
      \caption{All the 77 light curves during the transit of CoRoT-2b in front of its spotted host star. The gray solid line represent the model of a spotless star. The error bars of the data points indicate
 the rms of 0.0006.}
         \label{all}
   \end{figure}

A total of 77 transits were detected in the light curve with a high temporal resolution of 32 s, in a total of 134 days.  The rms of this signal was estimated from the out of transit data points to be $\sigma = 6\times 10^{-4}$ in relative flux units. Small light variations were detected during the transits, usually with fluxes between 3 and 10 sigma above the transit model without spots, the largest variation reaching 18 sigma. These intensity variations were interpreted as the signatures of star spots and were thus modeled. Figure~\ref{all} shows the light curves from all transits, where the vertical extent of each data point represents the rms of the signal. Also plotted on this figure is the model transit considering that no spots are present on the stellar surface (gray curves).

\section{The model} 

The physical characteristics of star spots is obtained by fitting the model described in \cite{silva03}. In the model adopted here, the star is a 2-D image with intensity decreasing following a quadratic limb darkening law according to \cite{alonso08}, whereas the planet is taken as a dark disc. The modeled light curve of the transit is obtained as follows. The planet position in its orbit is calculated every two minutes and the total flux is just the sum of all the pixels in the image (star plus dark planet). This yields the light curve, that is, the relative intensity as a function of time during the transit.
The model assumes that the orbit is circular, that is, null eccentricity (consistent with the measured eccentricity of $0.003 \pm 0.003$), and that the orbital plane is aligned with the star equator. In the case of CoRoT-2 the latter is a good assumption since, by measuring the Rossiter-McLaughlin effect, \cite{bouchy08} obtained that the angle between the stellar rotation axis and the normal of the orbital plane is only $7.2 \pm 4.5^o$.
The orbital parameters were taken from \cite{alonso08}, such as period of 1.743 day and orbital radius of 6.7 $R_{star}$.

The model also allows for the star to have features on its surface such as spots. The round spots are modeled by three parameters:
(i) intensity, as a function of stellar intensity at disk center, $I_c$ (maximum value); (ii) size, or radius, as a function of planet radius, $R_p$; and (iii) position: latitude (restricted to the transit path) and longitude. 
 On all the fitting runs, the latitude of the spots has remained fixed and equal to the planetary transit line, which for an inclination angle of $87.84^\circ$ is $-14.6^\circ$. This latitude was arbitrarily chosen to be South, thus the minus sign.  The longitude of the spot is measured with respect to the central meridian of the star, that coincides with the line-of-sight and the planet projection at transit center.

When the spot is near the limb, the effect of foreshortening is also featured in the model. However, this model does not account for faculae. Solar-like faculae have a negligible contrast close to the disc center, having significant contrast only close to the limb. Since the facular-to-spotted area ratio for CoRoT-2 is only Q=1.5 \citep{lanza09} instead of 9 as in the Sun, and we limit our analysis to spots between -70 and +70$^o$ from the central meridian, the photometric effect of faculae is very small and can be safely neglected.

The blocking of a spot by the planet during a transit implies an increase in the light detected, because a region darker than the stellar photosphere is being occulted. Thus the effect of many spots on the stellar surface is to decrease the depth of the transit, as can be seen in Figure~\ref{all}. This is turn will influence the determination of the planet radius, since a shallower transit depth results in a smaller estimate of the planet diameter when spots are ignored \citep{silva-valio10}. 

\begin{figure}
   \centering
  \includegraphics[width=8cm]{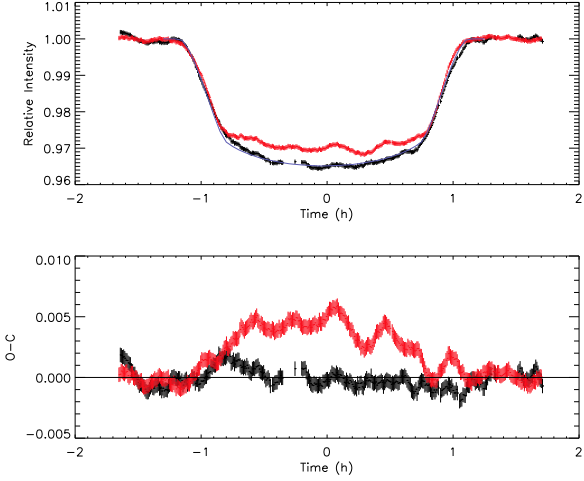}
      \caption{Top: Examples of transit light curves of CoRoT-2. The black crosses represent the 32nd transit assumed to occur when the star has minimum spot activity within the transit latitudes, whereas the gray data points represent a typical transit (the fifth one). The model light curve without any spots is shown as the thick solid curve. Bottom: Residual from the subtraction of the data minus the model for a star with no spots for the 5th (gray) and 32nd (black) transits. The vertical bars in both panels represent the estimated uncertainty of 0.0006 from out of transit data.               }
         \label{lc}
   \end{figure}

In this work, to estimate the best model parameters for the planet and its orbit, the deepest transit was sought.
Of all the 77 transits, the 32nd transit displays the lower  light variation (see Figure~\ref{all}). This was interpreted as the star having the minimum number of spots on its surface within the transit latitudes during the whole period of observation (134 days). The 32nd transit is shown in Figure~\ref{lc} as black crosses, for comparison, the fifth transit (gray crosses) is also shown  in the same figure. 

A light curve obtained from a model without any spot is shown as a thick solid curve on the figure. However, to obtain this model light curve it was necessary to use a planet radius of 0.172 stellar radius, instead of the 0.1667 stellar radius quoted on Table 1 of \cite{alonso08}, an increase of about 3\%. We note that this may not be a difference in the actual radius but rather an artifact due to the uneven spot coverage on the total surface of the star during that specific transit. Nevertheless, this implies that star spots can hinder the exact size estimate of a planet by making the transit light curve shallower than it would otherwise be \citep{silva-valio10}.

The actual radius of the planet may very well be 0.1667 $R_{star}$, since this was calculated from phase folded and averaged light curve. Supposing that the average area of the star covered by spots does not change during the whole period of observation (134 days), then when there are few spots along the transit line band ({\it e.g.} 32nd transit), there should be more spots on the remainder of the star. 

\subsection{Spot modeling}

\begin{figure}
   \centering
  \includegraphics[width=5cm]{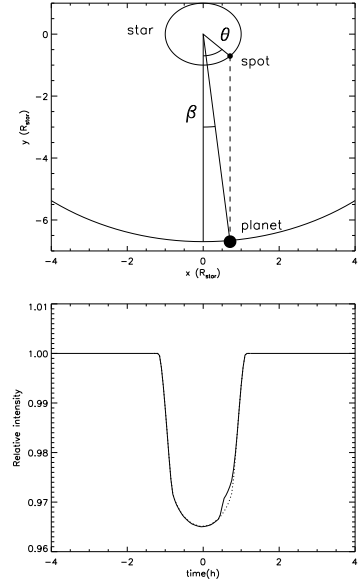}
      \caption{Top: Top view of the star and the planet in its orbit. A spot at 45$^o$ longitude on the stellar surface is depicted. Bottom: Light curve obtained from a star with a spot at 45$^o$ longitude. The dotted line represents a model for a transit in front of a spotless star.         }
         \label{diag1}
   \end{figure}

As mentioned in the previous section, the spots can be modeled by three basic parameters: intensity, radius, and longitude (the latitude is fixed at $-14.6^o$). All the fits were performed using the AMOEBA routine \citep{press92}. The longitude of the spot, is defined by the timing of the light variation within the transit. For example, the small ``bump" seen in the fifth transit (gray crosses in Figure~\ref{lc}) slightly to the left of the transit center, at approximately 0.1 h, is interpreted as being due to the presence of a spot at a longitude of $5.6^\circ$, where 0 longitude corresponds to the line-of-sight direction, taken as the central meridian of the star. According to the diagram shown in Figure~\ref{diag1}, the longitude of a spot may be estimated as:

\begin{equation} 
\theta = \sin^{-1} \left( \sin\beta {a \over R_s} \cos\alpha \right) \\
{\rm where} \quad  \beta = 2 \pi {{(t/24)} \over P_o}
\end{equation}

\noindent where $t$ is the time, measured with respect to the transit center, and given in hours, $P_o$ is the orbital period in days, and $\alpha$ is the latitude of the transit. The above equation is used to estimate the initial guess of the spots longitude, which was one of the parameters to be determined from the model fit to the data. 

The next step was to decide the maximum number of spots that were needed to fit each transit. Models with a maximum of 7, 8, and 9 spots on the stellar surface during each transit were tried. The results obtained from each run were qualitatively similar in all cases. It was found that in the case of 9 spots, the residuals were smaller than the uncertainty of the data (0.0006). Therefore, the results reported here are those from the fits with up to a total of nine different spots per transit. 

\section{Spot parameters resulting from the fits}

CoRoT-2 is a very active star, and many intensity variations were identified in each transit, implying that there are many spots present on the surface of the star at any given time. Beside its longitude, the spot signature also depends on its intensity and size. In fact, the flux perturbation from the spot is the product of the spot intensity and its area. Thus there is a degeneracy between the values of the radius and the intensity of one spot. 

Thus, to minimize the number of free parameters in the modeling, there are two alternatives. The first approach is to fix the radius of all the spots, for example, as half the planet radius, and fit each spot intensity and longitude for all transits. Another way is to fix the intensity of all the spots with respect to the maximum central intensity, $I_c$, and allow the spot radius and longitude to vary in each fit of the transit light curve. 

\begin{figure}
   \centering
  \includegraphics[width=8cm]{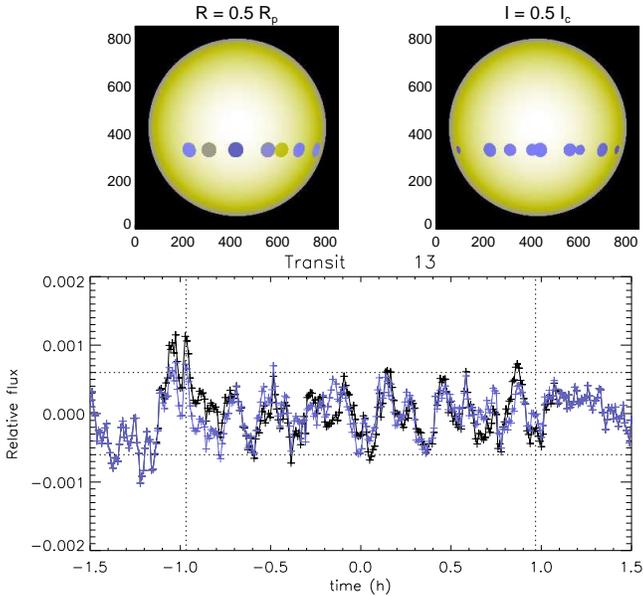}
      \caption{Top: Two examples of the model of the synthesized star with spots for  transit 13. The left panel shows spots with fixed radius of 0.5 $R_p$ and varying intensity, whereas the right panel presents spots with constant intensity of 0.5 $I_c$ and changing radius. Bottom: Residuals of the two models of constant radius (black) and intensity (red). }
         \label{ex}
   \end{figure}

Examples of the fitting by the two methods are shown in Figure~\ref{ex}. The two top panels represent the synthesized star with spots of: varying intensity and fixed radius of 0.5 $R_p$ (left) and varying radius and fixed intensity at 0.5 $I_c$  (right). The residuals of the data minus the two fits are shown in the bottom panel and show the little difference between the methods. 

The star was considered to have a varying number of spots (from 2 to a maximum of 9), with spot position defined at a certain longitude but a constant latitude of $-14.6^\circ$ (the transit latitude). The longitude considered is the topocentric longitude, that is, zero angle is defined as that of the line-of-sight, or transit center, when star, planet, and Earth are aligned.

\begin{figure}
   \centering
  \includegraphics[width=8cm]{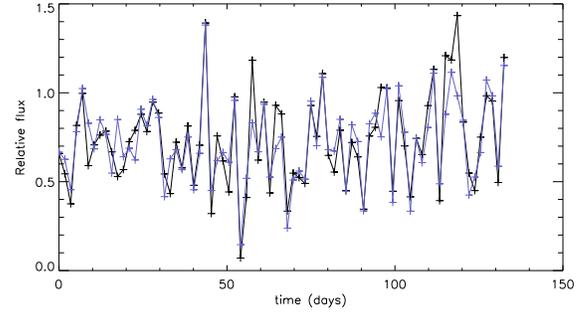}
      \caption{ Total relative flux of all spots per transit for both the model with constant spot intensity, 0.5 $I_c$ (black), and constant radius, 0.5 $R_p$ (gray).  }
         \label{flux}
   \end{figure}

Figure~\ref{flux} shows the total relative flux calculation, that is, the sum of all spot contrast times the squared radius (or area) in each transit, for both models. The spot contrast is taken as $1-f_i$, where $f_i$ is the relative intensity of the spot with respect to disc center intensity. As mentioned above, the flux, $F$, of a single spot is: $F \propto (1-f_i) R_{spot}^2$. For each transit the total relative flux was calculated by summing the flux of individual spots. As expected, the results from both models agree.

\subsection{Spot longitudes}

\begin{figure}
   \centering
  \includegraphics[width=8cm]{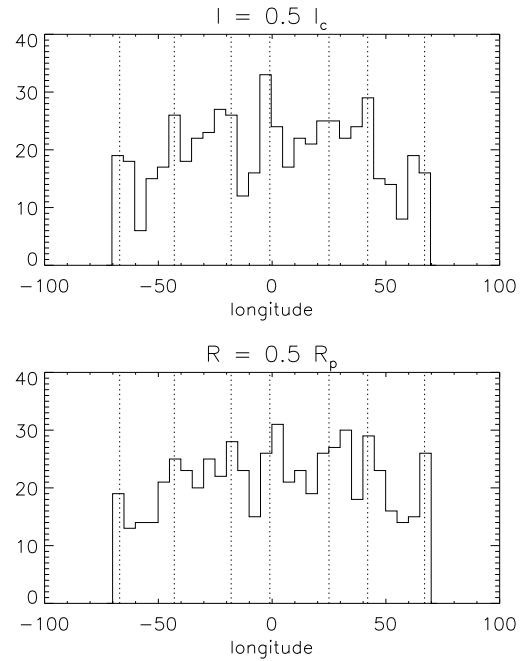}
      \caption{Histogram of spot longitudes for both models with constant intensity (top) and radius (bottom). The vertical dotted lines represent longitudes of -67, -43, -18, -1, 25, 42, and 67$^o$.}
         \label{long}
   \end{figure}

A histogram of the spots longitudes is shown in Figure~\ref{long} for the two models considered. Basically, seven main longitudes may be identified in the figure. Fortunately, they are approximately the same seven longitudes on both models, which is reassuring. The figure shows a slight predominance of the spots location at zero longitude, that is, the angle in the direction of the planet. 

These are topocentric longitudes, that is, they are not the ones located on the rotating frame of the star, but rather are measured with respect to an external reference frame. In order to obtain the longitudes in the stellar rotating frame, one needs an accurate period for the star. \cite{alonso08} report a $4.54$ days period, whereas \cite{lanza09} obtained a period of $4.52 \pm 0.14$ days in order to fit the rotational modulation of the out of transit data. A more precise estimate of the period is need in order to analyze the spot results. A detailed investigation of the rotational longitudes and the spot lifetime is underway and will be reported in an accompanying paper \citep{valio-lanza10}.

\subsection{Spot intensity and temperature}

\begin{figure}
   \centering
  \includegraphics[width=8cm]{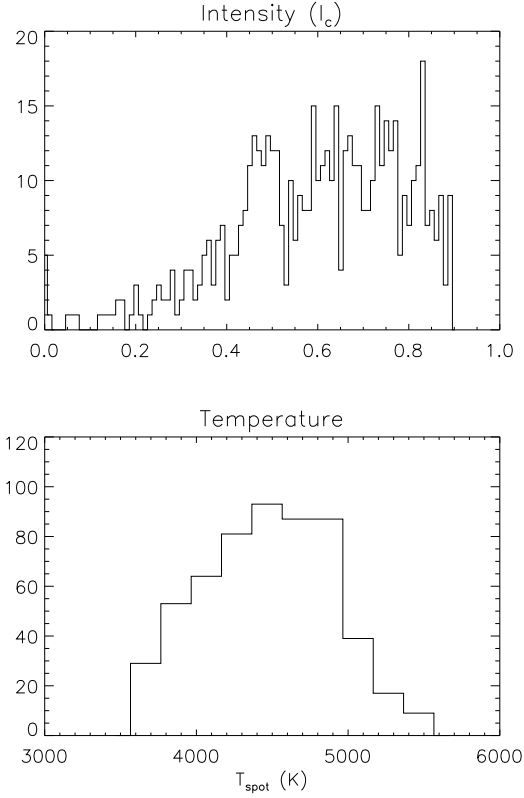}
      \caption{Results from the fit of the model with spots of fixed radius for all transits. Distributions of the spot intensity as a fraction of the central intensity (top), and spot temperature (bottom).              }
         \label{temp}
   \end{figure}

Spots with smaller intensity values, or high contrast spots, are spots cooler than those with intensity values close to $I_c$. The spot intensities obtained from the model with spots of fixed radius of $0.5 R_p$ are shown in the top panel of Figure~\ref{temp}. The figure shows that the spot intensities range from 0.4 to 0.9 $I_c$, with an average value of $0.60 \pm 0.19$ of the stellar central intensity.  This value is close to the value of $0.665 I_c$ used by \cite{lanza09}, that is the mean value of spots on the Sun.

These intensities can be converted to spot temperature by assuming blackbody emission for both the photosphere and the spots. The temperature is estimated as:

\begin{equation}
T_{spot} = {{h \nu} \over K_B} \left[ \ln \left( 1 + {{\exp^{\left( {{h \nu} \over {K_B T_{eff}}}\right)} - 1} \over f_i} \right) \right]^{-1}
\end{equation}

\noindent where $K_B$ and $h$ are Boltzmann and Planck constants, respectively, $\nu$ is the frequency associated to a wavelength of 600 nm, $f_i$ is the fraction of spot intensity with respect to the central stellar intensity, $I_c$, and $T_{eff}$ is the effective temperature of the star. Considering $T_{eff}=5625$ K \citep{alonso08}, the spots temperature range from 3600 to 5500 K, which are 100-2000 K cooler than the rest of the disk. The mean temperature of constant size spots on CoRoT-2 is $4600 \pm 400$ K.

\subsection{Spot radius}

\begin{figure}
   \centering
  \includegraphics[width=8cm]{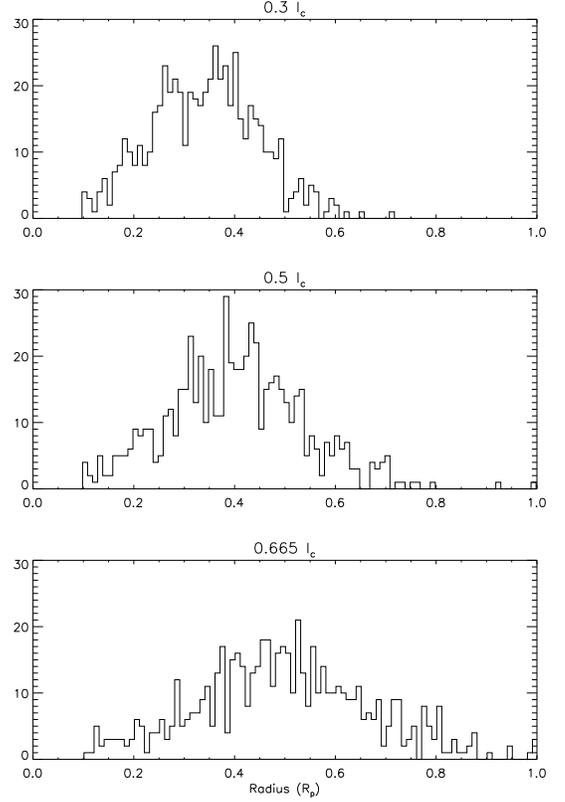}
      \caption{Results from the model fit for all transits: distributions of the spot radius in units of planet radius for three different fixed intensities as a fraction of the central intensity: 0.3 $I_c$ (top), 0.5 $I_c$  (middle), and  0.665 $I_c$(bottom).              }
         \label{rad}
   \end{figure}

 The distribution of spot radius from all the transits resulting from three simulations with different spot constant intensities: $0.3$, $0.5$, and $0.665$ of the central intensity, $I_c$ are shown on Figure~\ref{rad}. As can be seen from the histograms, as the spot constant intensity increases (or conversely its contrast decreases), the resulting radius also increases. This occurs in order to keep the spot flux the same. 

The models with fixed spot intensity at $0.3$, $0.5$, and $0.665 I_c$ resulted in spots with  average radius of $0.34 \pm 0.10\ R_p$, $0.41 \pm 0.13\ R_p$, and $0.50 \pm 0.17\ R_p$, respectively. The 0.5 $R_p$ value assumed in the spot model with fixed radius agrees with the average radius of spots with $0.665 I_c$, which interestingly is the mean intensity of sunspots \cite{lanza09}. 

The results show that the radius of the modeled spots varies from 0.2 to 0.7 $R_p$. Assuming a planet of $1.465 R_{Jup}$, this implies in spots with diameters of 40 to 150 Mm, with a mean value of about 100 Mm.

\subsection{Stellar surface area covered by spots}

\begin{figure}
   \centering
  \includegraphics[width=8cm]{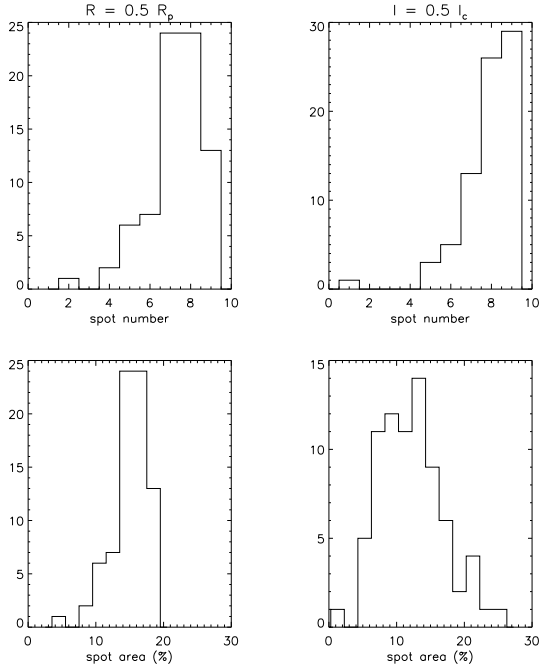}
      \caption{Top: Distribution of the number of spots detected on each planetary transit for both the constant radius, 0.5 $R_p$ (left), and intensity 0.5 $I_c$ (right) spot models. Bottom: Histograms of the stellar surface area covered by spots within the transit latitudes for each model (constant radius and intensity, left and right panels, respectively). }
         \label{area}
   \end{figure}

To estimate the area covered by spots on the surface of the star, only the significant spots should be taken into account. By significant we mean spot with contrast larger than 10\% of the stellar central intensity or radius larger than $0.1 R_p$, depending on the model.
According to this criterion, the number of spots on the surface of the star during each transit varied from 2 to 9 spots, with an average of 7 (for the constant radius model) or 8 (model with constant spot contrast) spots per transit. The transit with smaller number of spots corresponds to the 32nd transit discussed before. Histograms of the number of spots per transit are shown in the two top panels of Figure~\ref{area} for models with spots of constant radius (left) and intensity (right).

\begin{figure}
   \centering
  \includegraphics[width=6cm]{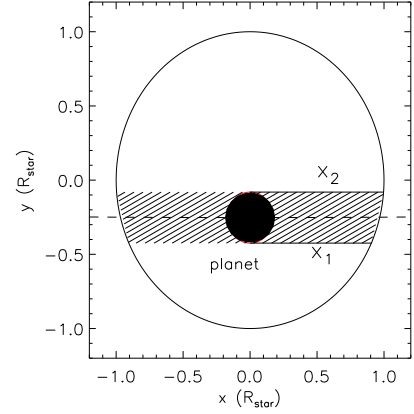}
      \caption{Diagram of the total area of the star occulted by the planet transit, represented by the hatched region.}
         \label{diag2}
   \end{figure}

The mean surface area covered by the spots during each transit is the sum of the area of all spots detected in that transit divided by the total area occulted by the passage of the planet (see the diagram on Figure~\ref{diag2}).
The total area, $A_{tot}$, of the transit band occulted by the planet is calculated as:

\begin{eqnarray}
A_{tot} = 2 R_p {7 \over 9} \pi \ {(x_1 + x_2) \over 2} R_s\\
{\rm where} \quad x_1 = \sqrt{1-\left[ \sin\alpha - {R_p \over R_s}\right]^2} \\
x_2 = \sqrt{ 1-\left[ \sin\alpha + {R_p \over R_s}\right]^2}
\end{eqnarray}

\noindent The $7/9$ factor arises because in the fit, only the spots between longitudes $-70^\circ$ and $+70^\circ$ are considered due to the difficulties in fitting spots too close to the limb where the light curve is very steep.

The surface area covered by spots on each transit, taking into account only the area of the transit band, of course, computed from the above equation is show in the bottom panels of Figure~\ref{area} for both models.
For the model with constant radius (bottom left), the spot surface area for each transit is the product of the number of spots on that transit multiplied by $\pi (0.5 R_p)^2$. On the other hand, for the model with constant spot intensity, the star surface area covered by spots is the sum of the area of all spots, that is, $ \Sigma \pi R_{spot}^2$, where $R_{spot}$ is the radius of the spot obtained from the fits.
The average values of the stellar surface area covered by spots are $16 \pm 3$\% and $13 \pm 5$\% for the models with constant radius ($0.5 R_p$) and intensity ($0.5 I_c$), respectively.

\begin{figure}
   \centering
  \includegraphics[width=6cm]{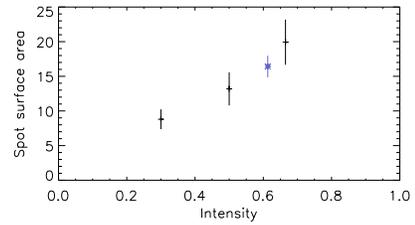}
      \caption{Mean stellar surface area covered by spots within the transit latitudes as a function of fixed spot intensity.  }
         \label{marea}
   \end{figure}

Because the data is only sensitive to the flux decrease due to the presence of spots, different values of spots intensity will produce fits with spots of different radius. The stellar area covered by spots was also calculated for the models with different intensities ($0.3$, $0.5$, and $0.665 I_c$) discussed in the previous subsection. 
A plot of the mean stellar area coverage as a function of the fixed spot intensity for each run is shown in Figure~\ref{marea}. 
Also plotted on the figure as an asterisk is the area coverage obtained from the model with fixed radius at $0.5 R_p$, where the mean intensity of $0.60 I_c$ was considered.

As can be seen from the figure, the area covered by spots within the transit latitudes increases as the spot intensity increases, varying, on average, between 10 and 20\%. This occurs because to fit the same variation in flux, a hotter spot (less contrast) needs to be larger, since the occulted area of the stellar surface is not so dark in this case. In summary, to account for the total decrease in light of the star due to the presence of hotter spots, one needs larger spots. This same trend was seen in the results of \cite{wolter09} from modelling of a single spot.

\section{Discussion and conclusions}

CoRoT-2 star is a young and reasonably active star. The presence of spots on the surface of the star can influence the determination of the orbital parameters of a planet in orbit by distorting the transit light curve in two ways \citep{silva-valio10}. One is the presence of spots on the limb of the star which will cause the transit duration to be shorter than it really is. 
The other distortion is to make the transit shallower if there are many spots on the surface of the star. This would cause the planet radius estimate to be smaller than its real value. The latter effect was observed in the dataset analyzed here, where the spot model applied here yield a radius of 0.172 $R_{star}$ instead of the 0.1667 $R_{star}$ listed in \cite{alonso08}. In this case, we do not believe that this represents a real difference in the planet radius, but rather an artifact of the spot distribution on the surface of the star at a given time.

Here the star was modeled as having up to 9 round spots at any given time on its surface at fixed positions (latitude and longitude) during the uninterrupted 134 days of observation by the CoRoT satellite. For each transit, the longitudes of the spots were obtained from a fit of the model to the data. The other free parameter obtained from the fit was either the spot radius or its intensity.

Two fitting approaches were performed on the transit light curves of CoRoT-2: the first one considered the radius of the spots to be fixed at 0.5 $R_p$, the second one kept the spot intensity at a constant value of the stellar central intensity, $I_c$. In the second approach, three different values of the fixed spot intensity were considered, one of them being the value of 0.665 $I_c$ used in \cite{lanza09}.
On every transit there were, on average, 7-8 spots on the visible stellar hemisphere within longitudes of $\pm 70^\circ$.

Despite the two methods with different fixed parameters of the spots, either radius or intensity, the results obtained for the spots characteristics were very similar. For example, the  longitudes are approximately the same and the spot surface area coverage are compatible. Also, the model with spots of fixed intensity at $0.665 I_c$ yields a mean value for the radius of $0.51 \pm 0.18 R_p$, agrees with the mean intensity of $0.60 \pm 0.19 I_c$ obtained from the model with spots of constant radius, $0.5 R_p$.  
This mean intensity is close to the value of $0.665\ I_c$ used by \cite{lanza09} which is the same as the sunspot bolometric contrast \citep{chapman94}. 
This agreement between the various approaches shows the robustness of the model. 

From the method of spots with fixed radius but varying intensities, the mean temperature of the spots was estimated as $4600 \pm 400$ K by considering blackbody emission for both the stellar photosphere and spots. These spots are about 1000 K cooler than the surrounding photosphere. On the other hand the runs of spots with varying radius but fixed intensity of 0.3, 0.5, and 0.665 $I_c$ yield mean radius of 0.35, 0.41, and 0.50 $R_p$. The increase in radius size means that darker spots (smaller intensity) are smaller than brighter ones (intensities close to $I_c$).

\cite{wolter09} modeled a single spot on one transit (here transit 54 of Figure~\ref{all}). The authors modeled the spot with different intensities, analogous to our procedure, and obtained a spot radius of $4.8^o$ (in degrees of stellar surface) for $0.3 I_c$. This specific feature, a ``bump" on the light curve transit was better fit by our model by two spots at longitudes of 3 and 18$^o$, with radius of $0.43$ and $0.39 R_p$ for the model of constant spot intensity of $0.3 I_c$. These radii are equivalent to $4.2$ and $3.9^o$, very similar to the results obtained by \cite{wolter09}. Moreover, \cite{wolter09} also confirm that spots with smaller contrast (higher intensity relative to the photosphere) need to be larger.

The spots on CoRoT-2, of the order of $\sim$100,000 km, are much larger than sunspots, about 10 times the size of a large sunspot (10 Mm). 
The mean surface area of the star covered by spots within the transit latitudes is about 10-20\%. This is larger than the 7 to 9\% of the total spotted area found by \cite{lanza09}. However, these values were estimated considering the whole star, that is, also the pole areas, where there are no spots in the case of the Sun. The values obtained here are only for the transit latitudes, which span approximately $20^\circ$ and are close to the equator. In this case, the latitudes coincide with he so called royal latitudes of the Sun, where most of the sunspots occurs.
It was reassuring to see that the results from both methods agreed very well with each other and specially with those of the out of transit data analysis \cite{lanza09}. 

Long term observations such as the one provided by CoRoT are paramount to understand the physics of stars spots. The model applied here to the CoRoT-2 data is capable of obtaining the physical properties of spots with the advantage of following their temporal evolution that is done in an accompanying paper \citep{valio-lanza10}. It will be very interesting to perform similar analyzes on data from other stars with planetary transits observed by the CoRoT satellite, especially for stars which are not solar-like.

\begin{acknowledgements}
We would like to thank everyone involved in the planning and operation of the CoRoT satellite which made these observations possible. A.S.V. acknowledges partial financial support from the Brazilian agency FAPESP (grant number 2006/50654-3).
\end{acknowledgements}

\end{document}